\documentclass[preprint,aps]{revtex4}
\usepackage{graphicx}
\usepackage{dcolumn}
\usepackage{bm}

\newcommand{\be}{\begin{equation}}
\newcommand{\ee}{\end{equation}}

\newcommand{\eqs}[2]{Eqs. (\ref{#1}) \& (\ref{#2})}

\newcommand{\eq}[1]{Eq. (\ref{#1})}

 \newcommand{\eqa}{\begin{eqnarray}}
\newcommand{\eeq}{\end{eqnarray}}  
 
\newcommand{\Eqsto}[2]{Equations (\ref{#1}) to (\ref{#2})} 
\newcommand{\sig}{\sigma_{\rm ex}}

\begin{document}

\title{Self-consistent interaction of neutrals and shocks in the local interstellar medium}

\author{S. Dastgeer\footnote{Electronic mail: {\tt dastgeer@ucr.edu}}, 
G. P. Zank and  N. Pogorelov }
\affiliation{Institute of Geophysics and Planetary Physics (IGPP),\\
University of California, Riverside, CA 92521. USA.\\
}

\begin{abstract}
The problem of hydrogen neutrals interacting with the heliospheric bow
shock region has received considerable attention recently, motivated
primarily by the hope that the Voyager spacecraft may soon encounter
the first of the heliospheric boundaries. The complexity of the charge
exchange interactions has limited our analytic understanding so far.
In this work we develop a semi-analytic one-dimensional model based
upon a double expansion technique to investigate the self-consistent
interaction of shock with neutrals.  The underlying method uses
the Boltzmann transport equation that describes neutral transport, and is
coupled to plasma protons predominantly through charge exchange
processes. The gyrotropic distribution function of the neutrals is
first expanded in terms of Legendre polynomials followed by associated
Laguerre harmonics. The resulting set of equations can be cast into a
square matrix whose eigenvalues depend upon sources and the plasma
distribution.  The interstellar protons, in a high $\beta$ limit, are
described by a hydrodynamic fluid with source terms that couple neutrals
and  plasma and further modify both  distributions in a
self-consistent manner.
\end{abstract}
\maketitle


\section{Introduction}
The interaction of the supersonic solar wind with the local
interstellar medium (LISM) is a problem of great topicality, as well
as promising fundamental insights for stellar systems interacting with
the interstellar envirnoment in general. The LISM is essentially a
partially ionized gas, and the coupling of atoms to the solar wind and
LISM plasma greatly complicates the interaction \cite{zank:1996}.
Such interactions have been been investigated using numerical models
such as a multifluid description where plasma and neutrals are treated
hydrodynamically \cite{zank:1995}, or by treating neutrals kinetically
and the plasma as an unmagnetized hydrofluid
\cite{hans:2000,baranov:1993}. The fluid description of neutrals in
the outer heliosphere can however be criticized because their mean
free paths are typically comparable to the size of the heliosphere.
The virtually collisionless neutrals therefore cannot equilibrate
thermally on heliospheric length-scales which essentially prevents the
use of a single fluid neutral description.

The modeling the solar wind-LISM interaction has been based almost
exclusively as coupled multi-dimensional models which include neutrals
more-or-less self-consistently. By contrast, very little analytical
work on the plasma-neutral coupling has accompanied the large-scale
modeling, and this is hampering our understanding of the very
complicated interaction. Here we develop a simple one-dimensional (1D)
transport equation for neutral atoms and investigate self-consistently
their influence on the structure and properties of hydrodynamic
shocks. Since the neutrals are coupled to the plasma via charge
exchange, we might expect the plasma to be cooled and the neutrals to
be heated downstream of the shock.  However some heated neutrals can
propagate back upstream, where they can undergo secondary charge
exchange and so heat the upstream unshocked fluid. Clearly one can
expect neutral atoms interacting with a shock wave to modify both the
structure and possibly the character of a shock. While obviously of
interest to the solar wind-LISM interaction, this work will have
interesting implications for any shock embedded in a partially ionized
medium, ranging from supernova shocks (SNR) to cometary bow shocks.

Beginning with the Boltzmann transport equation for the neutrals, we
employ a double expansion method to compute the neutral particle
distribution. The production and loss of neutrals are computed
similarly and couple to the plasma flow for which a hydrodynamical
description is adopted.

\section{Transport of neutrals}
The transport of neutrals across the plasma shock is governed by the
time dependent Boltzmann transport equation describing the evolution of
the neutral distribution function $f=f_H({\bf x}, {\bf v}, t)$ where
${\bf x}, {\bf v}, t$ denote the position, velocity and time
co-ordinates; thus
\be
\label{bol}
\frac{\partial f}{\partial t} + {\bf v} \cdot \nabla f + \frac{\bf F}{m}
\cdot \nabla_{\bf v} f = P({\bf x}, {\bf v}, t) - L({\bf x}, {\bf v}, t)
\ee
where $P$ and $L$ describe the production and loss of neutrals, mainly
due to charge exchange, and are given by \cite{zank:1995}
\[ P({\bf x}, {\bf v}, t)=f_p({\bf x}, {\bf v}, t)  \int f({\bf x}, {\bf v}', t)
|{\bf v}- {\bf v}'| \sig d^3{\bf v}';\]
\[  L({\bf x}, {\bf v}, t)=f({\bf x}, {\bf v}, t)  \int f({\bf x}, {\bf v}', t)
|{\bf v}- {\bf v}'| \sig d^3{\bf v}'.\] Here $f_p$ is the plasma
distribution function, ${\bf v}'$ is the speed of neutrals and $\sig$
is the charge exchange (between neutrals and plasma protons)
cross-section.  {\bf F} represents forces experienced by the neutrals
due to gravity and raditation pressure.  Let us now assume that the
neutral particles are gyrotropic and stationary in a frame of the
moving plasma fluid. The neutral speed can then be decomposed
according to $ {\bf v} = {\bf v}' + {\bf u}$, where ${\bf u} = u_p$ is
bulk flow velocity of the plasma. Upon using spherical co-ordinates,
the neutral velocity can be transformed as $ {\bf v} =
v\sin\theta\cos\phi
\hat{e}_x + v\sin\theta\sin\phi \hat{e}_y + v\cos\theta \hat{e}_z$. We
define  $\cos\theta = \mu$.  The  transport equation
describing the evolution of a neutral gyrotropic   distribution
function along the mean flow direction is given by
\eqa
\label{bol1}
\frac{\partial f}{\partial t} + 
\left( {\bf u} \cdot \nabla + \mu v \frac{\partial}{\partial z}\right)f  
+ \left[ \frac{\mu F_z}{mv} 
-\frac{\mu}{v} \left( \frac{\partial}{\partial t} + 
{\bf u}\cdot \nabla \right) u_z - \frac{1-\mu^2}{2} \nabla \cdot {\bf u} \right. \nonumber \\
\left. + \frac{1-3\mu^2}{2}\frac{\partial u_z}{\partial z} \right] v 
\frac{\partial f}{\partial v} 
+ \frac{1-\mu^2}{2} \left[ \frac{2 F_z}{mv} - \frac{2}{v} 
\left( \frac{\partial}{\partial t}+{\bf u}\cdot \nabla\right) u_z
+ \mu \nabla \cdot {\bf u} - 3\mu \frac{\partial u_z}{\partial z}\right]
\frac{\partial f}{\partial \mu} \nonumber \\
=  P({\bf x}, {\bf v}, t) - L({\bf x}, {\bf v}, t),
\eeq
where $u_z$ is the $\hat{z}$ component of the plasma velocity.  The
neutral transport Boltzmann \eq{bol1} is rather complicated by itself
and is coupled to the plasma evolution in two ways.  Firstly, the
spatial and temporal evolution of  plasma velocity modifies
the neutral distribution.  Secondly, the sources, i.e. the production
and the loss of neutrals by charge exchange, depend upon the relative
speed of the neutrals and the solar wind plasma proton species. The
absolute magnitude of this relative speed can be expressed in terms of
Legendre polynomials as $|{\bf v}- {\bf v}'| =\sum_{n=0}^\infty
a_n(v,v')P_n(\cos\xi)$, where the coefficients $a_n$ are to be
determined by orthogonality.  Using the addition theorem of spherical
harmonics for the orthogonal Legendre polynomials and carrying out the
angular integration between $0$ and $2\pi$ allows the production and
loss terms for the neutral species to be written as
\be
\label{pr}
 P({\bf x}, {\bf v}, t)= 2\pi f_p(v)\int_0^\infty
\int_{-1}^{+1} v'^2 f(v') \sig
\sum_{n=0}^\infty a_n(v,v') P_n(\mu)P_n(\mu') d\mu' dv'
\ee
\be
\label{lo}
 L({\bf x}, {\bf v}, t)= 2\pi f(v)\int_0^\infty 
\int_{-1}^{+1} v'^2 f_p(v') \sig
\sum_{n=0}^\infty a_n(v,v') P_n(\mu)P_n(\mu') d\mu' dv'
\ee
The charge exchange parameter $\sig$ has a logarithmically weak
dependence on the relative speed of the neutrals and plasma. It is
therefore assumed to be constant throughout our analysis.

\section{Solution by expansion}
The multidimensionality of the neutral distribution function,
described by \eq{bol1}, poses severe difficulties in its analytic
solution. We are therefore primarily concerned with reducing the
dimension of the \eq{bol1}. For this purpose, the analytic approach
developed by Zank et al \cite{zank:2000} has been used to reduce,
first, the $\mu$ dependence of the neutral particle distribution
function. The distribution function is expanded in terms of 
Legendre polynomials as $f( z, \mu,v, t) =
\sum_{n=0}^\infty (2n+1)P_n(\mu)f_n(z,v,t)$. With the help of the
orthogonality condition on the Legendre polynomials $\int_{-1}^{+1}
P_n(x)P_m(x) =
\delta_{m,n}/(2n+1)$, where $\delta_{m,n}$ is the Dirac delta function equals  unity for
$m=n$ and zero otherwise, the $n$th component of the evolution of the
neutral distribution function $f_n(z,v,t)$ is
\eqa
\label{bol2}
\frac{\partial}{\partial t} f_n + u_z\frac{\partial}{\partial z} f_n+
\left(\frac{n+1}{2n+1} \right)v\frac{\partial }{\partial z}f_{n+1}+
\frac{nv}{2n+1}\frac{\partial }{\partial z}f_{n-1}+
\frac{F_z}{m}\frac{1}{2n+1} \left[ n\frac{\partial }{\partial v}f_{n-1} \right. \nonumber \\
\left. +(n+1)\frac{\partial }{\partial v}f_{n+1} \right] 
-\frac{v}{2n+1} \left( \frac{u_z}{v}+1\right)\frac{\partial
u_z}{\partial z}
\left\{ \frac{n(n-1)}{2n-1}\frac{\partial }{\partial v}f_{n-2}
+\left[\frac{(n+1)^2}{2n+3}+   \right. \right. \nonumber \\ 
\left. \left. \frac{n^2}{2n-1}\right] \frac{\partial }{\partial v}f_{n} 
 +\frac{(n+1)(n+2)}{2n+3}\frac{\partial }{\partial v}f_{n+2}
\right\} 
+\frac{1}{2n+1}\left[(n+1)(n+2)f_{n+1} \right. \nonumber \\ 
\left.  -n(n-1)f_{n-1}\right]
\left(\frac{F_z}{mv}+\frac{u_z}{v}\frac{\partial u_z}{\partial z}\right)  
-\frac{\partial u_z}{\partial t} \left[ n\frac{\partial }{\partial
v}f_{n-1} +(n+1)\frac{\partial }{\partial
v}f_{n+1}+ \right. \nonumber \\  
\left. \frac{(n+1)(n+2)}{v}f_{n+1}  
 -\frac{n(n-1)}{v}f_{n-1}\right]  -\frac{1}{2n+1}
\left[ n(n+1) \left(\frac{n}{2n-1}-\frac{n+1}{2n+3} \right)f_n \right. \nonumber \\
\left. +\frac{(n+1)(n+2)(n+3)}{2n+3}f_{n+2} 
 -\frac{n(n-1)(n-2)}{2n-1}f_{n-2}
\right]\frac{\partial u_z}{\partial z}                            
=\frac{4N}{M\sqrt{\pi}}  \exp\left(-v^2\right) \nonumber \\  
\times \int_0^{\infty}dv' v'^2  \frac{a_n}{2n+1} f_n(v')    
-\frac{4N}{M\sqrt{\pi}} f_n(v)\int_0^{\infty}dv' v'^2
\exp\left(-v'^2\right) \frac{a_n}{2n+1}, 
\eeq
where $N=(n_{p_0}/n_{h_0})^{1/2}$ is a dimensionless parameter and
depends upon the ratio of equilibrium plasma ($n_{p_0}$) and neutral
($n_{h_0}$) densities, $M=u_{z_0}/v_{th}$ is plasma Mach number with
$u_{z_0}$ as upstream shock speed and $v_{th}$ is thermal velocity of
plasma protons. In the non-dimensional \eq{bol2}, we normalize the
distribution function, length-scales, time and velocity of neutrals
respectively as $\bar{f}=f/n_{p_0} \pi^{-3/2}/v_{th}^3, ~\bar{z}=z\sig
(n_{p_0} n_{h_0})^{1/2}, ~\bar{t}=t/v_{th} \sig (n_{p_0} n_{h_0})^{1/2},
~v=v'/v_{th}$. We have neglected the force terms assuming that gravity and
radiation pressure balance each other.
The rhs of \eq{bol2} represents source terms due to
constant charge exchange process and couples  the plasma to the neutral
dynamics. To work with this equation, the leading order harmonics,
i.e. $f_0, f_1, f_2, \cdots$, are good enough to describe a nearly
complete distribution of the neutral particles as they contribute
predominantly to the entire distribution function. Since the neutral
distribution function still depends upon position and velocity of the
particles which may vary in time, we choose another expansion to
eliminate the neutral velocity using $f_m(z,v,t)=\sum_{m=0}^\infty
\Theta_m(z,t) e^{-v}L_m^\mu(v)$ which doesn't lead to unphysically growing 
solution due to an exponentially damped weighting function in
$v$. This  results in a set of one-dimensional partial
differential equations for $\Theta_m(z,t)$ which can be cast into a
square matrix form of $n\times m$ order. The order of the matrix
depends mainly upon how many harmonics are  considered in the two
expansions. The general form of the matrix equations, however, looks
like 
 
\be
\left( \frac{\partial}{\partial t} + u_z \frac{\partial}{\partial z} \right)
\bar{\Lambda} + \bar{\Theta} \bar{\Lambda}=0 
\ee

\begin{displaymath}
{\rm with }~\bar{\Lambda}=
\left[\begin{array}{c}
\phi_0 \\
\psi_0 \\
\xi_0 \\
\phi_1 \\
\psi_1 \\
\xi_1 \\
\phi_2 \\
\psi_2 \\
\xi_2 \\
\vdots
\end{array}
\right]
,~~\bar{\Theta}= 
\left[ \begin{array}{ccccccccc}
c_{11} & c_{12} &  c_{13} & c_{14} & c_{15} & c_{16} & 0       & \ldots & d_1/M \\
c_{21} & c_{22} &  c_{23} & c_{24} & c_{25} & c_{26} & 0       & \ldots & d_2/M \\
c_{31} & c_{32} &  c_{33} & c_{34} & c_{35} & c_{36} & 0       & \ldots & d_3/M \\
0      & 0      &  0      & c_{44} & c_{45} & c_{46} &  c_{47} & \ldots & d_4/M \\
0      & 0      &  0      & c_{54} & c_{55} & c_{56} &  c_{57} & \ldots & d_5/M \\
0      & 0      &  0      & c_{64} & c_{65} & c_{66} &  c_{67} & \ldots & d_6/M \\
0      & 0      &  0      & 0      & 0      & 0      &  c_{77} & \ldots & d_7/M \\
\vdots & \vdots &  \vdots & \vdots & \vdots & \vdots &  c_{87} & \ldots & \vdots
\end{array} \right]
\end{displaymath}
where $\partial_z$ is partial derivative w.r.t the $z$ co-ordinate.
The coefficients ($C$'s, and $d$'s, whose actual expression are not
shown here) of the coupling matrix could in principle be a complicated
function of sources, the spatial and temporal evolution of the plasma
species and so too are the eigenvalues. Here $\phi_n,\psi_n,\xi_n,
\cdots$ are related to $f_1,f_2,f_3,\cdots$ by means of the Laguerre
polynomial expansion. The boundary condition (BC), which is crucial in
this problem, can be set appropriately using the two expansions. For a
typical problem, we inject Maxwellian neutrals, thermally equilibrated
with respect to the plasma, through the left boundary, while a
Dirichlet type-BC is set for the right boundary in the simulation. The
entire system of equations can be integrated numerically in the
presence of a high-$\beta$ ($\beta$ being the ratio of pressure and
magnetic energy of plasma particles) supersonic plasma for which we
adopt a hydrodynamical description, as discussed in the subsequent
section.

\section{Plasma dynamics}
The plasma protons can be described by the usual
compressible gasdynamical equations in which neutrals can influence
the evolution of plasma proton in a self-consistent manner.
\be
\label{density:SW}
\frac{\partial \rho}{\partial t} + \nabla \cdot (\rho {\bf u})=0,
\ee
\be
\label{mom:SW}
\frac{\partial}{\partial t} (\rho {\bf u})  + \nabla \cdot (\rho {\bf u} {\bf u})
+ \nabla p = {\bf Q}_m({\bf x},t), 
\ee
\be
\label{enr:SW}
\frac{\partial}{\partial t} \left( \frac{1}{2}\rho u^2 + 
\frac{p}{\gamma-1}\right) + \nabla \cdot \left(\frac{1}{2}\rho u^2  {\bf u} 
+ \frac{\gamma}{\gamma-1} p  {\bf u}  \right)
=Q_e({\bf x},t).
\ee
While the  plasma proton density is conserved,
i.e. \eq{density:SW}, the momentum and the energy,
\eqs{mom:SW}{enr:SW}, can be changed by the sources $Q_m$ and
$Q_e$ respectively.  The sources further couple the plasma fluid with
the neutral flow through the charge exchange processes. These can be
described by transfer integrals as computed below
\cite{hans:2000},
\[
{\bf Q}_m({\bf x},t) = \int \int ({\bf v}_H- {\bf v}_p)|{\bf v}_H- {\bf v}_p|\sig
 f_H({\bf x}, {\bf v}_H, t)  f_p({\bf x}, {\bf v}_p, t) 
 d^3{\bf v}_H d^3{\bf v}_p,
\]
\[
Q_e({\bf x},t) = \frac{1}{2}\int \int (v_H^2- v_p^2)|{\bf v}_H- {\bf v}_p|\sig
 f_H({\bf x}, {\bf v}_H, t)  f_p({\bf x}, {\bf v}_p, t) 
 d^3{\bf v}_H d^3{\bf v}_p.
\]
For a Maxwellian one-dimension plasma distribution, the above
 integrals reduce to the normalized form
\be
\label{qm} 
Q_m(z,t)= \frac{N}{M\pi^{3/2}} \int_{v'} 
\left( \frac{\sqrt{\pi}}{2}v'^2- v_{th} v'+ \frac{\sqrt{\pi}}{4}v_{th}^2\right)f(v') dv',
\ee
\be
\label{qe}
 Q_e(z,t)=\frac{N}{M\pi^{3/2}} \int_{v'}
\left[ \left( \frac{v'^3}{2}-\frac{v'}{4}v_{th}^2 \right)\sqrt{\pi}v_{th} -
\frac{v_{th}^2}{2} \left(v'^2+v_{th}^2\right)\right] f(v') dv',
\ee
where all the symbols have their usual meanings. \Eqsto{density:SW}{qe},
describing the evolution of the plasma protons,
along with the $n$th order matrix equation (for the neutrals), to form dynamically
a self-consistent problem.

\section{conclusion}
We have developed a self-consistent one-dimensional semi-analytic
model for transport of the neutrals governed by the time dependent
Boltzmann equation. The coupling of the neutrals to  plasma
protons, in a high $\beta$ collisionless interstellar flow, occurs
predominantly through the neutral-proton charge exchange. The latter
influences both the neutral and the plasma dynamics
significantly. A double-expansion method for the truncated neutral
Boltzmann equation was developed to seek a self-consistent solution of
the neutral-plasma coupled system.  The numerical simulation of the
entire system of equations is underway and will be presented
elsewhere.

\acknowledgments
 S. Dastgeer and G. P. Zank have been supported in part by NASA grants
NAG5-11621 and NAG5-10932 and an NSF grant ATM0296113. N. Pogorelov is supported by
NASA grant NAG5-12903.


\end{document}